%% file: DRDRCB_arxiv.tex
\documentclass{article}
\usepackage{spconf,amsmath,graphicx,cite,graphics,graphicx,amsmath,amssymb,epsfig}

\def\O{\mathcal{O}}
\include{mydefs}

\title{Data-Adaptive Reduced-Dimension Robust Beamforming Algorithms \vspace{-0.5em}}
\twoauthors
  {Samuel~D. Somasundaram\sthanks{This work was supported by MOD/DTIC under contract RT/COM/4/032.}, Nigel~H.~Parsons}
    {General Sonar Studies \\
    Thales Underwater Systems \\
    Cheadle Heath, Stockport, Cheshire, U.K.}
  {Peng~Li, Rodrigo~C.~de~Lamare}
    {Communications Research Group\\
    Department of Electronics\\
    University of York, U.K. \vspace{-0.5em}}
\begin{document}
\ninept

\maketitle
\begin{abstract}
We present low complexity, quickly converging robust adaptive
beamformers that combine robust Capon beamformer (RCB) methods and
data-adaptive Krylov subspace dimensionality reduction techniques.
We extend a recently proposed reduced-dimension RCB framework, which
ensures proper combination of RCBs with any form of dimensionality
reduction that can be expressed using a full-rank dimension reducing
transform, providing new results for data-adaptive dimensionality
reduction. We consider Krylov subspace methods computed with the
Powers-of-R (PoR) and Conjugate Gradient (CG) techniques,
illustrating how a fast CG-based algorithm can be formed by
beneficially exploiting that the CG-algorithm diagonalizes the
reduced-dimension covariance. Our simulations show the benefits of
the proposed approaches.

\end{abstract}
\begin{keywords}
Robust adaptive beamforming, dimensionality reduction, Krylov
subspace methods.
\end{keywords}
\section{Introduction and Preliminaries}
\label{sec:intro}

When implementing adaptive beamforming on arrays with large aperture
and many elements that operate in dynamic environments,
reduced-dimension techniques are often needed to speed-up the
convergence of beamforming algorithms and reduce the computational
complexity~\cite{SomasundaramRDRCBJournal}. This is of fundamental
importance in applications found in passive sonar and radar systems.
Furthermore, robust adaptive techniques are often required to
alleviate the deleterious effects of array steering vector (ASV)
mismatch, e.g., caused by calibration and pointing errors.
A popular class of these are the robust Capon beamformers (RCBs)
that exploit ellipsoidal, including spherical, uncertainty sets of
the ASV~\cite{VorobyovGL03,StoicaWL03,LiSW03,LorenzB05,LiS06}.
In~\cite{SomasundaramRDRCBConf,SomasundaramRDRCBJournal}, a
framework for combining reduced-dimension and RCB techniques was
derived, allowing rapidly converging, low complexity robust adaptive
reduced-dimension robust Capon beamformers (RDRCBs) to be formed. A
key contribution of that work was the derivation of a complex
propagation theorem that allows a reduced-dimension ellipsoid to be
derived from an element-space ellipsoid and any full-rank dimension
reducing transform (DRT). The reduced-dimension ellipsoid may then
be exploited by using an RCB in the reduced-dimension space.
In~\cite{SomasundaramRDRCBConf,SomasundaramRDRCBJournal}, only
data-independent dimensionality reduction was considered. Here, we
extend the framework developed in
\cite{SomasundaramRDRCBConf,SomasundaramRDRCBJournal} to
data-adaptive dimensionality reduction, providing new results useful
for exploiting a variety of scenarios that occur in practical
applications of robust beamforming algorithms.

The problem under consideration is the design of RDRCBs that are
suitable for large arrays. We consider Krylov subspace techniques
\cite{Honig02},\cite{GeKS06},\cite{deLamare08},\cite{Dietl01},\cite{Wang10}
for data-adaptive dimensionality reduction which are computed by the
Powers-of-R (PoR)
\cite{Honig02},\cite{GeKS06},\cite{deLamare08},\cite{delamarespl1},\cite{jidf}
and Conjugate-Gradient (CG) \cite{Wang10},{FLW10} algorithms. We
then develop RCB versions of the PoR and CG algorithms for large
arrays. We present a CG-based technique can exploit the fact that it
results in a diagonal reduced-dimension sample covariance matrix to
give particularly low-complexity data-adaptive beamforming
algorithms. Scenarios with large planar arrays are investigated
along with both non-degenerate ellipsoidal uncertainty and spherical
uncertainty sets.

In the following, $\E \left\{ \cdot \right\}$, $(\cdot)^T$,
$(\cdot)^H$, $(\cdot)^{-1}$ and $(\cdot)^\dagger$ denote the
expectation, transpose, Hermitian transpose, inverse and
Moore-Penrose pseudo-inverse operators, respectively.
Furthermore, $\left\| \cdot \right\|_2$, $\bN_\bX^l$, $\bPi_\bX$ and
$\bPi^\perp_\bX$ denote the two-norm, a basis for the left
null-space of $\bX$, the orthogonal projector onto the range space
of $\bX$ and the orthogonal projector onto the space perpendicular
to the range space of $\bX$, respectively.
Moreover, $\bX \geq 0$ or $\bX >0$ mean that the Hermitian matrix
$\bX$ is $+$ve semi-definite or $+$ve definite.

\subsection{Robust Capon Beamforming}
\label{sect:rcb}
We model the $k$th element-space array snapshot $\bx_k \in \mathbb{C}^M$ as
\begin{equation}
\bx_k = \ba_0 s_{0,k} + \bn_k,
\label{eq:snapmod}
\end{equation}
where $\ba_0$, $s_{0,k}$ and $\bn_k$ denote the true
signal-of-interest (SOI) ASV, the SOI complex amplitude and an
additive zero-mean complex Gaussian vector that incorporates the
noise and the interference. Assuming $s_{0,k}$ is zero mean and
uncorrelated with $\bn_k$, the array covariance matrix can be
written as $\bR_\bx = \E \left\{ \bx_k \bx_k^H \right\} = \sigma_0^2
\ba_0 \ba_0^H + \bQ_\bx$, where $\bR_\bx > 0$, $\sigma_0^2 = \E
\left\{ |s_{0,k}|^2 \right\}$ is the SOI power and $\bQ_\bx = \E
\left\{ \bn_k \bn_k^H \right\}$ is the noise plus interference
covariance matrix.
In practice, $\bR_\bx$ is often replaced by the sample covariance matrix (SCM)
\begin{equation}
\bRhat_{\bx} = \frac{1}{K} \sum_{k=1}^K \bx_k \bx_k^H,
\label{eq:sampcov}
\end{equation}
formed from $K$ snapshots.
In~\cite{StoicaWL03} (see, also~\cite{LiS06}), RCBs were derived by solving  $\max_{\sigma^2, \ba} \ \sigma^2 \ \text{s.t.} \ \bR_\bx - \sigma^2 \ba \ba^H \geq 0, \ \ba \in \mathcal{E}_M({\bar \ba}, \bE)$, which can be reduced to~\cite{StoicaWL03}
\begin{equation}
\min_{\ba} \ba^H \bR_\bx^{-1} \ba \quad  \text{s.t.} \quad \ba \in \mathcal{E}_M({\bar \ba}, \bE).
\label{eq:rcbform}
\end{equation}
The $M$-dimensional element-space ellipsoid $\mathcal{E}_M({\bar \ba}, \bE)$ is parameterized by ${\bar \ba}$, which often represents the assumed ASV, and $\bE \geq 0 \in \mathbb{C}^{M \times M}$, and can be written as
\begin{equation}
\mathcal{E}_M ({\bar \ba}, \bE) = \left\{ \ba \in \mathbb{C}^{M} \ \big| \ [\ba - {\bar \ba}]^H \bE [\ba - {\bar \ba}] \leq 1 \right\}.
\label{eq:nondegenrcb}
\end{equation}
For non-degenerate sets, $\bE>0$. To solve (\ref{eq:rcbform}), we assume that
\begin{equation}
{\bar \ba}^H \bE {\bar \ba} > 1
\label{eq:setcond}
\end{equation}
When $\bE=(1/\epsilon) \bI$, (\ref{eq:nondegenrcb}) reduces to a spherical uncertainty set, $\left\| \ba - {\bar \ba}\right\|_2^2 \leq \epsilon$, with radius $\sqrt{\epsilon}$ and (\ref{eq:setcond}) becomes $\left\|{\bar \ba} \right\|_2^2 > \epsilon$.
For non-degenerate ellipsoids, we can factor $\bE = \bE^{\frac{H}{2}} \bE^{\frac{1}{2}}$ and form $\breve{\ba} = \bE^{\frac{1}{2}} \ba$, $\breve{{\bar \ba}} = \bE^{\frac{1}{2}} {\bar \ba}$ and $\breve{\bR} = \bE^{\frac{1}{2}} \bR \bE^{\frac{H}{2}}$. Then, (\ref{eq:rcbform}) can be re-written using the following spherical constraint~\cite{LiSW03}
\begin{equation}
\min_{\breve{\ba}} {\breve \ba}^H {\breve \bR}^{-1} {\breve \ba} \ \text{s.t.} \ \left\| \breve{\ba} - \breve{{\bar \ba}} \right\|_2^2 \leq 1.
\label{eq:rcbspheremin}
\end{equation}
As shown in~\cite{LiSW03}, (\ref{eq:rcbspheremin}) can be solved via the eigenvalue decomposition (EVD) of ${\breve \bR}$, where computing the EVD is the most computationally expensive step. Denoting $\hat{\breve{\ba}}$ as the solution to (\ref{eq:rcbspheremin}), the solution to (\ref{eq:rcbform}) is formed as $\hat{\ba}_{0, \text{RCB}} = \bE^{-\frac{1}{2}} \hat{{\breve \ba}}$. The RCB power estimate is formed as ${\hat \sigma}_{0,\text{RCB}}^2 = \frac{\left\| {\hat \ba}_{0,\text{RCB}} \right\|_2^2/M}{{\hat \ba}_{0,\text{RCB}}^H \bR_\bx^{-1} {\hat \ba}_{0,\text{RCB}}}$
and the weight vector as ${\hat \bw}_{\text{RCB}} = \frac{\bR_\bx^{-1} {\hat \ba}_{0,\text{RCB}}}{{\hat \ba}_{0,\text{RCB}}^H\bR_\bx^{-1} {\hat \ba}_{0,\text{RCB}}}$.

\section{ Robust Capon Beamforming Framework with Data-Adaptive Reduced-Dimension}
\label{sect:rdrcb} In reduced-dimension methods, the $k$th
element-space snapshot, $\bx_k \in \mathbb{C}^{M }$, is projected
onto an $N$-dimensional subspace (with $N<M$) using a DRT $\bD \in
\mathbb{C}^{M\times N}$, yielding the reduced-dimension snapshot,
$\by_k = \bD^H \bx_k$, where $\by_k \in \mathbb{C}^{N}$.
As shown in~\cite{SomasundaramRDRCBJournal,SomasundaramRDRCBConf}, this leads to the following RDRCB problem
$\max_{\sigma^2, \bb} \ \sigma^2 \ \text{s.t.} \ \bR_\by - \sigma^2 \bb \bb^H \geq 0, \ \bb \in \mathcal{E}_N({\bar \bb}, \bF)$,
where $\bb = \bD^H \ba$, $\bR_\by = \bD^H \bR_\bx \bD$ and $\mathcal{E}_N({\bar \bb}, \bF)$ denote the reduced-dimension ASV, covariance and uncertainty ellipsoid, respectively, which can be reduced to
\begin{equation}
\min_{\bb} \bb^H \bR_\by^{-1} \bb \quad  \text{s.t.} \quad \bb \in \mathcal{E}_N({\bar \bb}, \bF).
\label{eq:rdrcbform}
\end{equation}
The following theorem is used to derive $\mathcal{E}_N({\bar \bb}, \bF)$.

{\em Propagation Theorem:}~\cite{SomasundaramRDRCBConf,SomasundaramRDRCBJournal} The propagation of the element-space ellipsoid (\ref{eq:nondegenrcb}), with $\bE \geq 0 \in \mathbb{C}^{M \times M}$, through the mapping $\bD^H \ba - \bI_N \bb = {\bf 0}$, where $\bD \in \mathbb{C}^{M \times N}$ has full column rank, yields the ellipsoid $\mathcal{E}_N({\bar \bb},\bF)$ [see (\ref{eq:nondegenrcb})] with
\begin{eqnarray}
{\bar \bb} &=& \bD^H {\bar \ba} \label{eq:bbar} \\
\bF &=& \bD^\dagger(\bE - \bE \bN_\bD^l [(\bN_\bD^l)^H \bE \bN_\bD^l]^\dagger (\bN_\bD^l)^H \bE) (\bD^\dagger)^H.
\label{eq:bF}
\end{eqnarray}

For data-adaptive dimensionality reduction, ${\bar \bb}$ and $\bF$
need updating each time the DRT is updated. If we use (\ref{eq:bF})
for updating, then we observe that $\bN_\bD^l$, $[(\bN_\bD^l)^H \bE
\bN_\bD^l]^\dagger$ and $\bD^\dagger$ need calculating, which are
expensive operations.
Fortunately, if the original element-space ellipsoid is non-degenerate, such that $\bE >0$, we can simplify (\ref{eq:bF}). Then, $[(\bN_\bD^l)^H \bE \bN_\bD^l]^\dagger = [(\bN_\bD^l)^H \bE \bN_\bD^l]^{-1}$ and we can write
\begin{eqnarray}
\bF 
&=& \bD^\dagger \bE^{\frac{1}{2}} \bPi^\perp_{\bE^{\frac{1}{2}} \bN_\bD^l} \bE^{\frac{1}{2}} (\bD^\dagger)^H \notag \\
&=& \bD^\dagger \bE^{\frac{1}{2}} \bPi_{\bE^{-\frac{1}{2}} \bD} \bE^{\frac{1}{2}} (\bD^\dagger)^H \notag \\
%
%
&=& \left[ \bD^H \bE^{-1} \bD \right]^{-1},
\label{eq:bFNDE}
\end{eqnarray}
where $\bPi^\perp_{\bE^{\frac{1}{2}} \bN_\bD^l} = \bI - \bE^{\frac{1}{2}} \bN_\bD^l [(\bN_\bD^l)^H \bE \bN_\bD^l]^{-1} (\bN_\bD^l)^H \bE^{\frac{1}{2}}$.
The inverse $\bE^{-1}$ can be computed offline and therefore, the online computation of $\bF$ reduces to the computation of an $N \times N$ inverse. Note that, in general, we will need to compute $\bF^{\frac{1}{2}}$, $\bF^{\frac{H}{2}}$ and $\bF^{-\frac{1}{2}}$ [see Section~\ref{sect:rcb}], which can all be obtained from the EVD of $\left[ \bD^H \bE^{-1} \bD \right]$. Note also that, in general, we will require the EVD of $\breve{\bR}_\by = \bF^{\frac{1}{2}} \bR_\by \bF^{\frac{H}{2}} = \bF^{\frac{1}{2}} \bD^H \bR_\bx \bD \bF^{\frac{H}{2}}$.
Thus, in general, two $N$-dimensional EVDs will be required, one decomposing $\breve{\bR}_\by$ and one decomposing $\left[ \bD^H \bE^{-1} \bD \right]$.
When the element-space uncertainty set is a sphere, so that in (\ref{eq:nondegenrcb}), $\bE = \frac{1}{\epsilon} \bI$, then
\begin{equation}
\bF = \frac{1}{\epsilon} (\bD^H \bD)^{-1}.
\label{eq:bF1}
\end{equation}
In this case, if the DRT is orthogonal, $\bF$ in (\ref{eq:bF1}) reduces to $\bF = \frac{1}{\epsilon} (\bD^H \bD)^{-1} = \frac{1}{\epsilon} \bI_N$.
Thus, if the element-space set is a sphere and the DRT is orthogonal, then $\bF$ can be written analytically and only one EVD is required.
Denoting ${\hat \bb}_0$ as the solution to (\ref{eq:rdrcbform}), we form the RDRCB weight vector as
\begin{equation}
{\hat \bw}_{\text{RDRCB}} = \frac{\bR_\by^{-1} {\hat \bb}_0}{{\hat \bb}_0^H\bR_\by^{-1} {\hat \bb}_0}.
\label{eq:wrdrcb}
\end{equation}
The weight vector (\ref{eq:wrdrcb}) operates on the reduced-dimension data. The weight vector that operates on the original element-space data is given by ${\hat \bw}_{\text{RDRCB,ES}} = \bD {\hat \bw}_{\text{RDRCB}}$. An estimate of $\ba_0$ can be formed as ${\hat \ba}_{0} = (\bD^H)^\dagger {\hat \bb}_0 = \bD(\bD^H\bD)^{-1}{\hat \bb}_0$.
Note that if ${\hat \bb}_0 = \bD^H {\hat \ba}_0$, then ${\hat \ba}_0
= \bPi_\bD {\hat \ba}_0$, where $\bPi_\bD$ is an orthogonal
projection matrix onto the column space of $\bD$.
Given ${\hat \ba}_0$, we form the RDRCB SOI power estimate as
\begin{equation}
{\hat \sigma}_{0,\text{RDRCB}}^2 = \frac{(\left\| {\hat \ba}_0 \right\|_2^2/M)}{{\hat \bb}_0^H \bR_\by^{-1} {\hat \bb}_0} = \frac{{\hat \bb}_0^H (\bD^H \bD)^{-1} {\hat \bb}_0}{M {\hat \bb}_0^H \bR_\by^{-1} {\hat \bb}_0}.
\label{eq:RDRCBpower}
\end{equation}
%

\section{Data-Dependent Dimension Reduction}
\label{sect:dddr}

Here, we consider Krylov methods that use the PoR and CG algorithms
to compute the matrix that performs dimension reduction.

\subsection{PoR (Non-Orthogonal) Krylov Basis}

The standard PoR method for creating a Krylov DRT is to form
\begin{equation}
\bD = \left[\begin{array}{cccc}\frac{\bar \ba}{\left\|{\bar \ba}  \right\|_2} & \frac{\bRhat_\bx {\bar \ba}}{\left\| \bRhat_\bx {\bar \ba} \right\|_2}& \hdots & \frac{\bRhat_\bx^{N-1} {\bar \ba}}{\left\| \bRhat_\bx^{N-1} {\bar \ba} \right\|_2} \end{array} \right],
\end{equation}
which can be formed iteratively. That is, starting with $\bkappa_1 =
{\bar \ba}$, and $\bD_1 =\frac{\bar \ba}{\left\|{\bar \ba}
\right\|_2}$, for $i = 2,\hdots,N$, calculate
\begin{equation}\bkappa_i = \bRhat_\bx \bkappa_{i-1},\end{equation}
\begin{equation}\bd_i = \frac{\bkappa_i}{\left\| \bkappa_i \right\|_2}\end{equation}
and
\begin{equation}\bD_i
= \left[\begin{array}{cc} \bD_{i-1} & \bd_i \end{array}
\right]\end{equation}.
The cost of calculating $\bkappa_i$ from $\bkappa_{i-1}$ is $\O(M^2)$ and calculating $\bd_i$ is $\O(M)$ . Thus, calculating the Krylov DRT costs $\O(NM[M+1])$ flops.
The resulting Krylov DRT is non-orthogonal (NO) and therefore, to compute the NO-Krylov RDRCB, two $N$-dimensional EVDs will need computing, even if the original element-space set is spherical.


\subsection{PoR Orthogonal Krylov Basis}

In~\cite{GeKS06}, the PoR orthogonal Krylov (O-Krylov) subspace
technique was proposed and suggested for applications where the
model order is highly variable and time-varying.
To form the O-Krylov DRT, let $\bkappa_1 = {\bar \ba}$, $\bD_1 =
{\bar \ba}/ \left\| {\bar \ba} \right\|_2$, and for $i =
2,\hdots,N$, calculate
\begin{equation}\bkappa_i = \bPi_{\bD_{i-1}}^\perp \bRhat_\bx
\bkappa_{i-1}\end{equation},
\begin{equation}
\bd_i =
\frac{\bkappa_i}{\left\| \bkappa_i \right\|_2}
\end{equation}
and
\begin{equation}
\bD_i = \left[\begin{array}{cc} \bD_{i-1} & \bd_i \end{array}
\right]
\end{equation},
where $\bPi_{\bD_{i-1}}^\perp = \bI - \sum_{k=1}^{i-1} \bd_k
\bd_k^H$ can be updated efficiently in $\O(M^2)$ operations using
$\bPi_{\bD_i}^\perp = \bPi_{\bD_{i-1}}^\perp - \bd_i \bd_i^H$. Given
$\bPi_{\bD_{i-1}}^\perp$, updating $\kappa_i$ and $\bd_i$ costs
$\O(2M^2)$ and $\O(M)$. Thus, the calculation of one new column of
$\bD$ costs $\O(3M^2 + M)$, so that calculation of the
\mbox{O-Krylov} DRT costs $\O(NM[3M+1])$, which is roughly three
times more expensive than calculating the standard \mbox{NO-Krylov}
DRT.
Since the resulting DRT is orthogonal, as discussed earlier, for spherical uncertainty sets only one EVD is required to compute the RDRCB.


\subsection{Conjugate Gradient Method}

Using the approach outlined in~\cite{Dietl01}, the CG DRT can be
formed by setting, $\bd_1 = {\bar \ba}$, $\br_1 = -{\bar \ba}$, and
then for $i = 2,\hdots, N$, update using
\begin{equation}
\alpha_i = -\frac{\bd_i^H \br_i}{\bd_i^H \bRhat_\bx \bd_i},
\end{equation}
\begin{equation}
\br_{i+1} = \br_i + \alpha_i \bRhat_\bx \bd_i,
\end{equation}
\begin{equation}
\beta_i = \frac{\bd_i^H \bRhat_\bx \br_{i+1}}{\bd_i^H \bRhat_\bx
\bd_i} \end{equation}
and
\begin{equation}
\bd_{i+1} = - \br_{i+1} + \beta_i \bd_i .
\end{equation}
The cost of computing $\bRhat_\bx \bd_i$ is $\O(M^2)$. Given $\bRhat_\bx \bd_i$, the cost of computing $\alpha_i$ is $\O(2M)$. Updating $\br_{i+1}$ is $\O(M)$. The cost of computing $\beta_i$, given $\bRhat_\bx \bd_i$ and the denominator of $\alpha_i$ is $\O(M)$. Then, updating $\bd_{i+1}$ is $\O(M)$. Thus, the total cost to compute a new column of the CG DRT is $\O(M^2 + 5M)$. Thus, the total cost to calculate the CG DRT is $\O(NM[M + 5])$, which is almost the same as calculating the NO Krylov DRT.
Since the CG DRT is NO, we would expect that we would need two EVDs to compute the CG-RDRCB. However, in the next section, we illustrate how a fast CG-based RDRCB can be obtained by exploiting that the CG DRT diagonalizes the SCM so that
\begin{equation}
\bRhat_\by = \bD^H \bRhat_\bx \bD = \bLambda_{\text{CG}},
\label{eq:CGRy}
\end{equation}
where $\bLambda_{\text{CG}}$ is a diagonal matrix and $\bD = [ \bd_{1} \ldots \bd_{N}]$ is the DRT matrix. 
%

\section{Fast Conjugate-Gradient RDRCB}
Here, we illustrate how only one $N$-dimensional EVD is required to solve the CG-RDRCB under either spherical or non-degenerate uncertainty.
In general, we will be solving
\begin{equation}
\min_\bb \bb^H \bR_\by^{-1} \bb \ \text{s.t.} \ \left[\bb - {\bar \bb} \right]^H \bF \left[\bb - {\bar \bb} \right] \leq 1.
\label{eq:NDRDRCBmin}
\end{equation}
Usually, at this stage one would transform with $\bF^{\frac{1}{2}}$ to give a spherical uncertainty set. However, from (\ref{eq:CGRy}), we observe that $\bR_\by^{-1} = \bLambda_{\text{CG}}^{-1}$, so that (\ref{eq:NDRDRCBmin}) can be written as
\begin{equation}
\min_\bb \bb^H \bLambda_{\text{CG}}^{-1} \bb \ \text{s.t.} \ \left[\bb - {\bar \bb} \right]^H \bF \left[\bb - {\bar \bb} \right] \leq 1.
\label{eq:CGRDRCBmin}
\end{equation}
Letting $\bM = \bLambda_{\text{CG}}^{-\frac{1}{2}} \bD^H \bE^{-1} \bD \bLambda_{\text{CG}}^{-\frac{H}{2}}$, $\check{\bb} = \bLambda_{\text{CG}}^{-\frac{1}{2}} \bb$ and $\check{{\bar \bb}} = \bLambda_{\text{CG}}^{-\frac{1}{2}} {\bar \bb}$, we can rewrite (\ref{eq:CGRDRCBmin}) as
\begin{equation}
\min_{\check{\bb}} \check{\bb}^H \check{\bb} \ \text{s.t.} \ \left[\check{\bb} - \check{{\bar \bb}} \right]^H \bM^{-1} \left[\check{\bb} - \check{{\bar \bb}} \right] \leq 1.
\label{eq:TransCGRDRCBmin}
\end{equation}
We form the Lagrangian using the real Lagrange multiplier $\mu$
\begin{equation}
L(\check{\bb}, \mu) = \check{\bb}^H \check{\bb} + \mu \left( \left[\check{\bb} - \check{{\bar \bb}} \right]^H  \bM^{-1} \left[ \check{\bb} - \check{{\bar \bb}} \right] - 1 \right).
\end{equation}
Setting $\frac{\partial L(\check{\bb}, \mu)}{\partial \check{\bb}^H} = {\bf 0}$ yields
\begin{equation}
\hat{{\check \bb}} = \left( \frac{\bM}{\mu} + \bI \right)^{-1} \check{{\bar \bb}} = \check{{\bar \bb}} - \left[ \mu \bM^{-1} + \bI \right]^{-1} \check{{\bar \bb}},
\label{eq:CGhatbchk}
\end{equation}
where we have used the matrix inversion lemma to obtain the term after the second equality.
Using (\ref{eq:CGhatbchk}) in the constraint equation in (\ref{eq:TransCGRDRCBmin}) yields
\begin{equation}
h(\hat{{\check \bb}}, \mu) = \check{{\bar \bb}}^H \left[ \mu \bM^{-1} + \bI \right]^{-1} \bM^{-1} \left[ \mu \bM^{-1} + \bI \right]^{-1} \check{{\bar \bb}}^H.
\label{eq:CGLmu}
\end{equation}
Letting $\bM = \bU \bLambda \bU^H$ denote the EVD of $\bM$, where $\bLambda = \text{diag} \left\{[ \begin{array}{ccc} \lambda_1 & \hdots & \lambda_N \end{array} ]\right\}$ is a diagonal matrix containing the eigenvalues in non-increasing order on its main diagonal and $\bU$ contains the associated eigenvectors, we can write (\ref{eq:CGLmu}) as
\begin{equation}
h(\hat{{\check \bb}}, \mu) = \sum_{n=1}^N \frac{\lambda_n |c_n|^2}{\left(\mu + \lambda_n \right)^2},
\end{equation}
where $c_n$ is the $n$th element of $\bc = \bU^H \check{{\bar \bb}}$.
Since we can write $\bM=\bM^{\frac{1}{2}} \bM^{\frac{H}{2}}$, where $\bM^{\frac{1}{2}} = \bLambda_{\text{CG}}^{-\frac{1}{2}} \bD^H \bE^{-\frac{1}{2}}$, we know that $\bM$ is non-negative definite~\cite{Strang88,StoicaM05} and therefore, it has non-negative eigenvalues. Thus, $h(\hat{{\check \bb}}, \mu)$ is a monotonically decreasing function of $\mu > 0$.
For $\mu = 0$, we obtain
\begin{eqnarray}
h(\hat{{\check \bb}}, 0) =  \check{{\bar \bb}}^H \bM^{-1}  \check{{\bar \bb}} = {\bar \bb}^H \left[\bD^H \bE^{-1} \bD \right]^{-1} {\bar \bb} = {\bar \bb}^H \bF {\bar \bb}.
\end{eqnarray}
Note that, to exclude a non-trivial solution we require that, ${\bar \bb}^H \bF {\bar \bb} > 1$. Since we require $h(\hat{{\check \bb}}, \mu) = 1$, it is clear that $\mu \neq 0$. Further, it is clear that $\lim_{\mu \rightarrow \infty} h(\hat{{\check \bb}}, \mu) = 0$, therefore, there is a unique solution $\mu > 0$ to $h(\hat{{\check \bb}}, \mu)=1$, which can be found, e.g., by Newton search.
Once $\mu$ has been found, $\hat{{\check \bb}}$ is found using (\ref{eq:CGhatbchk}) and the solution to (\ref{eq:CGRDRCBmin}) is formed as $\hat{\bb}_0 = \bLambda_{\text{CG}}^{\frac{1}{2}} \hat{{\check \bb}}$.
We can use $\hat{\bb}_0$ and $\bR_\by^{-1} = \Lambda_{\text{CG}}^{-1}$ in (\ref{eq:wrdrcb}) to form the adaptive weights. To form the power estimate using (\ref{eq:RDRCBpower}), we need to evaluate $\left[ \bD^H \bD \right]^{-1}$. If the uncertainty set is spherical, then we can evaluate this quantity from the EVD of $\bM$ and $\bLambda_{\text{CG}}$, which are already available. For a general, non-degenerate ellipsoid this quantity will need computing.

\begin{figure}[t!]
\centerline{
\begin{tabular}{c}
\includegraphics[width=21pc] {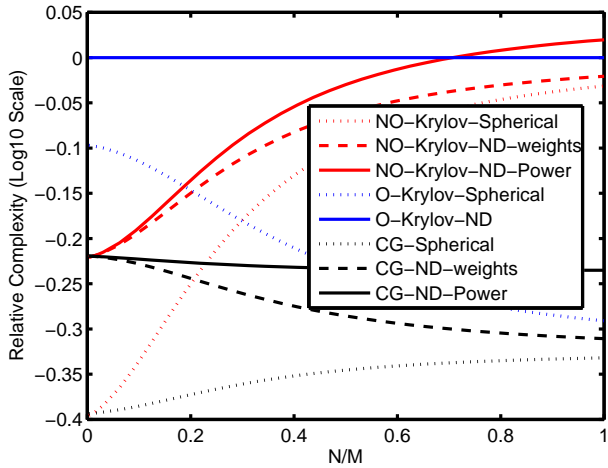}
\end{tabular}
} \vspace{-1em} \caption{Relative complexities of different
data-dependent RDRCBs.} \label{fig:complexity}
\end{figure}

Fig.~\ref{fig:complexity} shows the relative complexities as $N$ is increased from 1 to $M$, for $M = 320$, illustrating that the CG-based algorithms are significantly cheaper than the other methods.

%

\section{Numerical Examples}
\label{sect:numex}

\begin{figure}[t!]
\centerline{
\begin{tabular}{c}
\includegraphics[width=21pc] {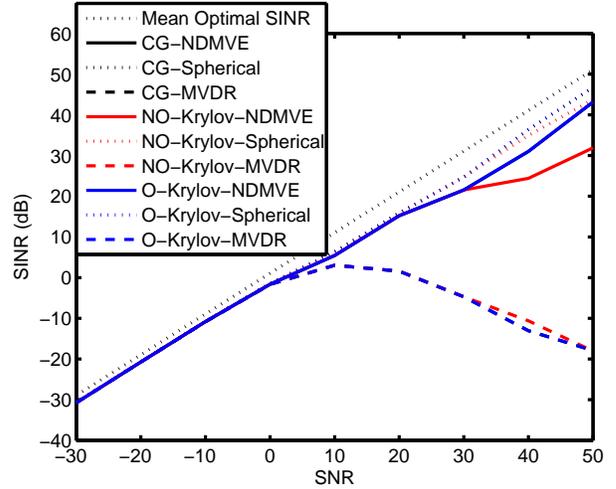}
\end{tabular}
} \vspace{-1em} \caption{SINR versus SNR for the algorithms
analyzed.} \label{fig:SINRvsSNR}
\end{figure}

In this section, we assess the performance of the proposed
algorithms through numerical examples. For an $M=320$,
$\lambda/2$-spaced planar array with $M_h = 40$ elements in a row
and $M_v=8$ rows, we simulated data with covariance matrix $\bR_\bx
= \sigma_0^2 \ba_0 \ba_0^H + \bQ_\bx$, with $\bQ_\bx = \sum_{i=1}^d
\sigma_i^2 \ba_i \ba_i^H + \sigma_s^2 \bI + \sigma_{\text{iso}}^2
\bQ_{\text{iso}}$, where $\bQ_\bx$ consists of terms due to $d$
zero-mean uncorrelated interferences, where for the $i$th interferer
$\sigma_i^2$ and $\ba_i$ denote the source power and ASV, a term
modeling sensor noise $\sigma_s^2 \bI$, with sensor noise power
$\sigma_s^2$, and a term modeling an isotropic ambient noise
$\sigma_{\text{iso}}^2 \bQ_{\text{iso}}$, with power
$\sigma_{\text{iso}}^2$. The isotropic noise covariance is given by
$[\bQ_{\text{iso}}]_{m,n} = \text{sinc} [\pi g_{mn}]$, where
$g_{mn}$ is the distance between the $m$th and $n$th sensors in
units of wavelength.
The $i$th source (SOI or interference) ASV is simulated according to
$\ba_i = \ba({\bar \btheta}_i + \delta_i) + \sigma_{e,i} \be_i$,
where $\be_i$ is a zero-mean complex circularly symmetric random
vector with unit norm. When $\delta_i \neq 0$ an AOA error exists
and when $\sigma_{e,i} \neq 0$, an arbitrary error exists.
We assume azimuth and elevation beams spaced at $1/M_h$ and $1/M_v$
in cosine space and, using the methods described
in~\cite{SomasundaramJP12}, design tight-spherical uncertainty sets
and non-degenerate minimum volume ellipsoidal (NDMVE) sets  based on
the expected AOA errors given the spacing of the beams.
Fig.~\ref{fig:SINRvsSNR} shows SINR versus SNR for....??? The
results show that the CG and O-Krylov results are the same, whilst
the NO-Krylov results diverge for very high SNRs. It is clear that
the robust RDRCB version exploiting spherical or non-degenerate
NDMVE sets, provide much better robustness at high SNRs compared to
the standard MVDR-based implementations.

 \vspace{-0.5em}
%

%
%


\bibliographystyle{IEEEbib}
\bibliography{sam_extra}

\end{document}

%% file: mydefs.tex
\def \b1{{\bf 1}}

\def\ba{{\bf a}}
\def\bb{{\bf b}}
\def\bc{{\bf c}}
\def\bd{{\bf d}}
\def\be{{\bf e}}

\def\bn{{\bf n}}

\def\br{{\bf r}}

\def\bw{{\bf w}}
\def\bx{{\bf x}}

\def\by{{\bf y}}

\def\bD{{\bf D}}

\def\E{{E}}
\def\bE{{\bf E}}

\def\bF{{\bf F}}

\def\bI{{\bf I}}

\def\bM{{\bf M}}
\def\bN{{\bf N}}

\def\bQ{{\bf Q}}

\def\bR{{\bf R}}

\def\bRhat{{\hat \bR}}

\def\bU{{\bf U}}

\def\bX{{\bf X}}

\def\bkappa{{\boldsymbol \kappa}}

\def\btheta{{\boldsymbol \theta}}

\def\bLambda{{\boldsymbol \Lambda}}
\def\bPi{{\boldsymbol \Pi}}

